# Circular Rectifiction of 3D Video and Efficient Modification of 3D-HEVC

Jarosław Samelak
Inst. of Multimedia Telecommunications
Poznań University of Technology
Poznań, Poland,
jaroslaw.samelak@gmail.com

Marek Domański
Inst. of Multimedia Telecommunications
Poznań University of Technology
Poznań, Poland,
marek.domanski@put.poznan.pl

*Abstract*— **Video acquired from multiple cameras located along a line is often rectified to video virtually obtained from cameras with ideally parallel optical axes collocated on a single plane and principal points on a line. Such an approach simplifies video processing including depth estimation and compression. Nowadays, for many application video, like virtual reality or virtual navigation, the content is often acquired by cameras located nearly on a circle or on a part of that. Therefore, we introduce new operation of circular rectification that results in multiview video virtually obtained from cameras located on an ideal arc and with optical axes that are collocated on a single plane and they intersect in a single point. For the circularly rectified video, depth estimation and compression are simplified. The standard 3D-HEVC codec was designed for rectified video and its efficiency is limited for video acquired from cameras located on an arc. Therefore, we developed a 3-D HEVC codec modified in order to compress efficiently circularly rectified video. The experiments demonstrate its better performance than for the standard 3D-HEVC codec.**

## I. Introduction

At the beginning of the second decade of 2000s, extensive efforts were aimed at development of multiview and 3D video coding technology for the content acquired from multiple cameras densely distributed on a line. The research resulted in development of multiview and 3D extensions of Advanced Video Coding (AVC) [4] and High Efficiency Video Coding (HEVC) [5] like MV-HEVC (Multi-View HEVC) and 3D-HEVC [6]. This development was related to expected applications of autostereoscopic displays that simultaneously display several dozens of views related to the locations slightly shifted along a straight line. Unfortunately, autostereoscopic displays still have not gained sufficient popularity until now. This is one of the reasons why multiview and 3D profiles of AVC and HEVC are of limited usage now.

More recently, rash development of the virtual reality technology caused raising interest in compression of multiview or 3D video acquired by cameras located around a scene, often nearly on a circle or on an arc [8-11]. Unfortunately, both MV-HEVC and 3D-HEVC were designed the video content acquired from camera densely distributed on a line and for such content they provide substantial bitrate reduction as compared to simulcast HEVC [6, 7]. Unfortunately this gain reduces to very small or even negligible values for the content obtained from cameras sparsely distributed on a circle or even arbitrarily located around a scene [1, 2]. This effect results from the usage of simple disparity-compensated inter-view predictions in MV-HEVC and 3D-HEVC. In the references, this problem [3] was dealt by modifications of 3D-HEVC codecs by the use of real 3D mappings for inter-view predictions instead of the simple disparity-compensated prediction [1, 2]. Unfortunately, the usage of real 3D mappings results in substantial increase of the computational effort for inter-view predictions.

In 3D-HEVC, it is assumed that input video was acquired with cameras located ideally on a line with their optical axes being parallel on a single plain [6]. Such requirement is impossible to meet in practice due to differences between cameras and difficulties in positioning them ideally on a line. Therefore, prior to compression, multiview video is usually rectified, which corresponds to correcting positions of cameras and suppressing the results of differences in their properties (Fig. 1). In multiview and 3D video processing, video acquired from multiple camera located along a line is usually rectified in order to process video virtually obtained from cameras with ideally parallel optical axes collocated on a single plane [12, 13]. It should be stressed that rectification does not correct real positions of cameras, but transforms the views obtained from the cameras onto a common plane, which is tantamount to acquiring video with ideally positioned cameras. Such an approach simplifies video processing including depth estimation and compression.

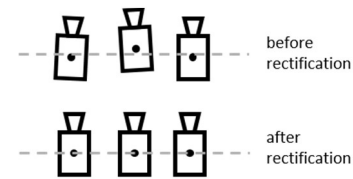

Fig. 1. Linear camera setup before and after rectification.

Here, we propose to perform similar operation for multiview/3D video obtained from cameras located on an arc. Obviously, they are never located on an ideal arc. Therefore, in this paper, we introduce new operation of circular rectification that results in multiview video virtually obtained from cameras located on an ideal arc and with optical axes that are collocated on a single plane and are intersecting in a single point. In the paper, we describe the procedure for circular rectification of multiview video. For the circularly rectified video, depth estimation and compression are simplified.

The research was supported by the Ministry of Education and Science of the Republic of Poland under the subvention for research SBAD.
The work of M. Domański was partially done on leave at Nagoya University, Japan under support of Foundation for Polish Science.

As mentioned before, the 3D-HEVC codec was developed and tested predominantly for rectified video and its efficiency is limited for video acquired from cameras located on an arc. Therefore, in this paper, we develop a 3-D HEVC codec modified in order to compress efficiently circularly rectified video. The experiments demonstrate its better performance than for the standard 3D-HEVC codec.

Let recapitulate, for this paper, the two main goals are:
- develop the concept and the procedure for circular rectification,
- propose an efficient modification of 3D-HEVC codec for processing of circularly rectified 3D video.

## II. MAIN IDEA

Our main idea is to introduce a similar approach as used for applications of 3D-HEVC, but for cameras arranged ideally on a circle, with their optical axes lying on a plane and directed towards centre of the circle (Fig. 2). Setting up such multi-camera system in practice would be even more challenging, however it provides much more information about the scene, which is crucial for many applications.

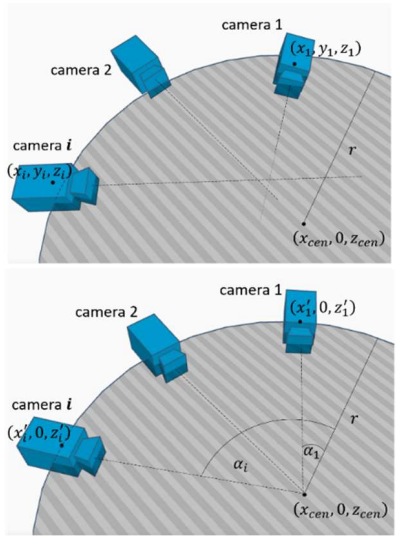

Fig. 2. Circular camera setup before (top) and after proposed rectification (bottom).

The proposed approach constitutes an alternative processing path to two aproaches already described in the references and depicted by the two left pathes in Fig. 3. The first step of the proposal is circular rectification, i.e. correction of real camera positions to points located on a circle, with cameras' optical axes parallel to the ground and directed towards the centre of the circle, and transformation of the input 3D video according to the change.

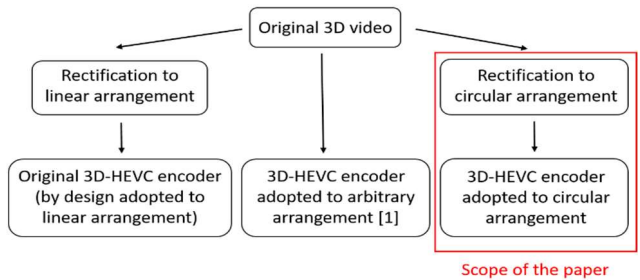

Fig. 3. Available and proposed paths of processing and encoding of 3D video.

## III. CIRCULAR RECTIFICATION

In this section, the proposed process of circular rectification is escribed, including derivation of circle parameters, modification of camera parameters and transforming test sequences.

### A. Camera parameters

In the paper, it is assumed that all camera parameters are known and represented by intrinsic parameter matrix [3×3] $\mathbb{K}$ (1), rotation matrix [3×3] $\mathbb{R}$ and 3-component translation vector $\mathbb{T}$. The derivation of the camera parameters is described elsewhere [14 – 16].

$$\mathbb{K} = \begin{bmatrix} f_x & c & o_x \\ 0 & f_y & o_y \\ 0 & 0 & 1 \end{bmatrix}, \quad (1)$$

where: $f_x, f_y$ – focal lengths, $o_x, o_y$ – coordinates of the optical centre, $c$ – skew factor.

The abovementioned intrinsic and extrinsic camera parameters can be used to calculate the projection matrix [4×4] $\mathbb{P}$ for each camera using Eq. (2),

$$\mathbb{P} = \begin{bmatrix} \mathbb{K} & 0 \\ 0^T & 1 \end{bmatrix} \begin{bmatrix} \mathbb{R} & \mathbb{T} \\ 0^T & 1 \end{bmatrix}. \quad (2)$$

Then, the positions of corresponding points in two camera views $A$ and $B$ can be derived according to Formula (3) that can be used to transform the video by "moving" the location of the camera (matrix $\mathbb{P}$ is usually nonsingular).

$$\begin{bmatrix} z_B \cdot x_B \\ z_B \cdot y_B \\ z_B \\ 1 \end{bmatrix} = \mathbb{P}_B \cdot \mathbb{P}_A^{-1} \begin{bmatrix} z_A \cdot x_A \\ z_A \cdot y_A \\ z_A \\ 1 \end{bmatrix}, \quad (3)$$

where: $(x_A, y_A)$, $(x_B, y_B)$ are the positions of corresponding points in view $A$ and $B$, respectively,
$z_A, z_B$ are the depth values of corresponding points in view $A$ and $B$, respectively,
$\mathbb{P}_A, \mathbb{P}_B$ are the projection matrices for view $A$ and $B$.
It is also assumed that the cameras are located roughly on a circle around the scene, otherwise rectification could introduce inacceptable distortions to the input video. Similar multi-camera systems have already been successfully set up and used to record 3D video test sequences (e.g. in [9, 22]).

### B. Derivation of circle parameters and new camera positions

The first step of circular rectification is finding the parameters of the circle, based on the positions (represented by translation vectors $\mathbb{T}$) of the cameras. The circle is represented by position of its centre ($x_{cen}$, 0, $z_{cen}$) and radius $r$. In order to find the aforementioned parameters, we use circle Equation (4) with positions ($x_i$, 0, $z_i$) of each of the $N$ cameras, and perform non-linear regression by minimizing the sum of squares $S$ according to Equation (5).

$$(x_i - x_{cen})^2 + (z_i - z_{cen})^2 = r^2 \quad (4)$$

$$S = \sum_{i=1}^{N} (\sqrt{(x_i - x_{cen})^2 + (z_i - z_{cen})^2} - r)^2 \quad (5)$$

It should be noted that vertical positions are ignored ($y_{cen} = 0$, $y_i = 0$) because proposed rectification assumes that all cameras, as well as the centre of the circle, are located at the same height.

After derivation of circle parameters, the next step is to find for each camera its modified position ($x_i'$, 0, $z_i'$) on the circle, the closest to the original location. Figure 4 presents both original and new camera positions of one of real 3D video test sequences.

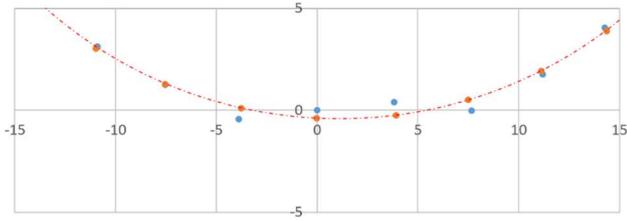

Fig. 4. Top view of a multi-camera system with original camera positions (blue dots) and shifted to ideal cirle (orange dots) for Breakdancers test sequence.

*C. Rotation of cameras into ideally circular arrangement*

In the proposed process of circular rectification the goal is not only to correct the locations of the cameras, but also to direct their optical axes precisely towards the centre of a circle derived in the previous subsection. To achieve that, modification of rotation matrices is necessary.

Rotation matrix $\mathbb{R}$ represents combined rotation of a camera around 3 orthogonal axes. In the ideally circular arrangement, optical axes are assumed to be on a single plane (parallel to the ground). Such camera rotation can be represented by the following matrix (6):

$$\mathbb{R}'_i = \begin{bmatrix} \cos\alpha_i & 0 & \sin\alpha_i \\ 0 & 1 & 0 \\ -\sin\alpha_i & 0 & \cos\alpha_i \end{bmatrix}, \quad (6)$$

where $\alpha_i$ is the angle between position of $i$-th camera and circle centre, therefore:

$$\cos\alpha_i = \frac{z'_i - z_{cen}}{r}, \quad \sin\alpha_i = \frac{x'_i - x_{cen}}{r} \quad (7)$$

All the necessary parameters: the circle centre position ($x_{cen}$, $z_{cen}$) and its radius $r$, as well as the modified $i$-th camera position ($x_i'$, $z_i'$) are already derived, thus there is no need to provide additional input parameters to find the rotation matrix of rectified cameras.

*D. Modification of intrinsic camera parameters*

Previous subsections *B* and *C* present process of deriving extrinsic parameters of cameras located on an ideal circle. This subsection shortly describes how to evaluate internal camera parameters of circularly rectified sequence.

First, the skew factor is set to *c=0*, similarly to linear rectification in the state-of-the-art 3D-HEVC [7, 17]. Then, focal length $f_y$ and vertical component of principal point $o_y$ are averaged and set equal for every camera. For comparison, in 3D-HEVC the values of $f_y$ and $o_y$ were not used at all. A more sophisticated approach is required to derive the horizontal component of principal point coordinate $o_x$. The cameras in a quasi-circular arrangement are usually directed towards the centre of recorded scene, which can be (and often is) much closer to the cameras than the centre of a circle. Therefore, due to modification of rotation matrices towards the centre of a circle, the field of view of each camera can be significantly modified. This can result in only small proportion of original field of view being covered by given camera after circular rectification (Fig. 5). In such a case, rectified views would contain only a small part of original content, which is highly unwanted.

Shifting camera field of view can be achieved not only by rotating the camera, but also by changing its principal point (Fig. 6). In the proposed circular rectification technique, new principal $o_x'$ is calculated for each camera to assert roughly the coverage of recorded scene as without rectification. It is done by projecting point equal to original optical centre $o_x$ onto 3D space. New value of $o_x'$ should compensate rectification of rotation matrix, thus projecting it onto 3D space should result in the same location.

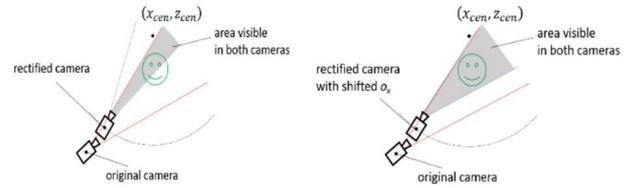

Fig. 5. Problem with shifted field of view after circular rectification

Fig. 6. Rectified camera directed towards circle centre and with modified optical centre $o_x'$ to shift its field of view

## IV. INTER-VIEW PREDICTION FOR CIRCULARLY RECTIFIED VIDEO AND MODIFIED HEVC CODEC

As mentioned before, inter-view prediction in 3D-HEVC is simplified due video rectification. On the other hand, inter-view prediction that uses full perspective projection requires complex operations on matrices and is noticeably slower. The proposed circular rectification is a trade-off between the abovementioned approaches. On one hand, the cameras can be located on a circle, which results in better coverage of recorded scene. On the other hand, the number of camera parameters required to describe the system is significantly reduced after circular rectification. The authors observed that rectification of camera parameters may be used to optimize the inter-view prediction for faster compression of circularly rectified 3D video. After applying rectified intrinsic and extrinsic camera parameters derived in Section III for full projection equations (3), the authors derived simplified formulas for projecting points between views *A* and *B*:

$$z_B = (x_A - o_{xA})\frac{z_A}{f_x}\sin\Delta\alpha + (z_A - r)\cos\Delta\alpha + r \quad (8)$$

$$y_B = o_y + \frac{z_A}{z_B}(y_A - o_y) \quad (9)$$

$$x_B = o_{xB} + \frac{1}{z_B}[(x_A - o_{xA})z_A\cos\Delta\alpha - (z_A - r)f_x\sin\Delta\alpha] \quad (10)$$

where $\Delta\alpha = \alpha_B - \alpha_A$.

The above formulas allow to predict the position of a point in view *B* based on its position in view *A* and circular camera parameters. The authors have modified the inter-view prediction in 3D-HEVC by replacing standard disparity derivation with point projection that uses the above equations. Moreover, a number of prediction techniques that exploit inter-view similarities have been modified, e.g. Inter-view Motion Prediction, View Synthesis Prediction, Neighboring Block Disparity Vector, Depth-oriented Neighboring Block Disparity Vector, Illumination Compensation. Table I compares parameters used by state-of-the-art 3D-HEVC and modified 3D-HEVC encoders for compression of circular and arbitrary 3D video. One may observe that rectified circular camera setup requires much less parameters than arbitrary, and 2 more values compared to unmodified 3D-HEVC. In the proposed encoder, all parameters, including non-standard $o_y$, $\alpha$, $r$, are transmitted in the bitstream in Video Parameter Set (VPS).

Obviously, the changes in the bitstream result in the proposed codec not being compliant with the 3D-HEVC

standard. Nevertheless, the authors prove that support for rectified circular 3D video compression could be added with only minor changes in the bitstream syntax.

TABLE I. CAMERA PARAMETERS NEEDED IN THE CASE OF LINEAR, CIRCULAR AND ARBITRARY CAMERA LOCATIONS

| Parameter name | Arbitrary camera setup | Circular camera setup | Linear camera setup |
|---|---|---|---|
| Horizontal focal length | $f_x$ | $f_x$ | $f_x$ |
| Vertical focal length | $f_y$ | - | - |
| Horizontal optical centre | $o_x$ | $o_x$ | $o_x$ |
| Vertical optical centre | $o_y$ | $o_y$ | - |
| Skew factor | $c$ | - | - |
| Translation | $\mathbb{T} = [t_x, t_y, t_z]$ | $\alpha, r$ | $t_x$ |
| Rotation | $\mathbb{R} = [r_{11}, ..., r_{33}]$ | | - |

## V. METHODOLOGY OF EXPERIMENTS

The goal of the experiments is to assess the compression efficiency and encoding time using the aforementioned codecs with respect to standard 3D-HEVC codec. Additionally, authors compared the encoding time of intra-view prediction only, for both modified 3D-HEVC encoders (circular and arbitrary). Compression efficiency was compared by measuring average bitrate reduction for luma component of texture views, using Bjøntegaard metric [20].

The experiments were conducted by encoding 7 views of 4 commonly-used multiview test sequences [21 - 23]. Encoding was performed at 4 QP values (25, 30, 35, 40) and 100 frames. The test sequences were rectified by full perspective projection (3). Obviously, rectified view may contain some unfilled areas – these are interpolated from surrounding content. Both texture and depth maps were rectified. Encoders were configured identically, following Common Test Conditions for 3D video experiments [18]. The only difference was in input camera parameters, which were prepared according to requirements of every encoder (Table I). Circular rectification of test sequences was done in pre-processing phase, so it does not affect encoding time results. Moreover, all three encoders were based on the same version of 3D-HEVC state-of-the-art, publicly available test model HTM-13.0 [19].

## VI. EXPERIMENTAL RESULTS

Table II shows bitrate reduction for the modified 3D-HEVC encoder for compression of circularly rectified video against unmodified 3D-HEVC and modified 3D-HEVC for arbitrary camera locations (from [1]). Our proposal reduces bitrate on average by 6% compared to the state-of-the-art technique. This is because 3D-HEVC does not perform accurate inter-view prediction if video was acquired by camera arrangements other than linear. Compared to encoder that supports any camera setup, solution optimized for circular arrangement provides slightly better results. The difference is caused by a lower number of camera parameters required by the proposal and simpler inter-view prediction, which results in reduced numbers of errors.

Table III presents reduction of total encoding time, while Table IV compares inter-view prediction time between two modified encoders. It should be noted that the proposed encoder is up to 10% faster than encoder with full perspective projection, and at the same time it's inter-view prediction is 44 times faster, due to much simpler projection equations, optimized for circularly rectified 3D video. Surprisingly, the proposed encoder was also faster than plain 3D-HEVC by roughly 4%, even though inter-view prediction of the former is more complex than state-of-the-art technique. The reason of such phenomenon is related to results for compression efficiency (Table II). As mentioned before, modified 3D-HEVC is more accurate in predicting content of circularly rectified video.

TABLE II. BITRATE REDUCTION (NEGATIVE VALUES) FOR THE PROPOSED 3D-HEVC FOR CIRCULARLY RECTIFIED VIDEO.

| Sequence | vs 3D-HEVC | vs 3D-HEVC modified for arbitrary arrangements [1] |
|---|---|---|
| Ballet | -8.14% | -0.23% |
| Breakdancers | -8.22% | 0.82% |
| BBB_Flowers | -3.15% | -0.54% |
| Poznan_Blocks | -4.61% | -0.23% |
| **Average** | **-6.03%** | **-0.05%** |

TABLE III. ENCODING TIME REDUCTION (NEGATIVE VALUES) FOR THE PROPOSED 3D-HEVC FOR CIRCULARLY RECTIFIED VIDEO.

| Sequence | vs 3D-HEVC | vs 3D-HEVC modified for arbitrary arrangements [1] |
|---|---|---|
| Ballet | -1.89% | -10.21% |
| Breakdancers | -8.50% | -7.46% |
| BBB_Flowers | -1.78% | -7.66% |
| Poznan_Blocks | -4.50% | -9.89% |
| **Average** | **-4.17%** | **-8.81%** |

TABLE IV. REDUCTION OF THE INTER-VIEW PREDICTION TIME DUE TO CIRCULAR RECTIFICATION APPLIED

| Sequence | vs 3D-HEVC modified for arbitrary arrangements [1] |
|---|---|
| Ballet | -97.96% |
| Breakdancers | -97.79% |
| BBB_Flowers | -97.77% |
| Poznan_Blocks | -97.43% |
| **Average** | **-97.74%** |

## VII. CONCLUSIONS

In the paper, the authors propose a novel approach to 3D video compression. 3D video acquired by cameras located nearly on an arc is proposed to undergo circular rectification proposed and discussed in Section III. The state-of-the-art 3D-HEVC technique is modified for efficient compression of such video. The codec modification is mostly related to modification of the inter-view prediction. The total bitrate for the modified codec appears to be lower by about 6% as compared to the standard 3-D HEVC applied for the standard MPEG test 3D video sequences captured by cameras distributed on an arc. The authors developed a process for correction of camera parameters to ideal circle together with circular video rectification. Moreover, projection equations for optimized inter-view prediction of circularly rectified 3D video were derived and implemented on top of 3D-HEVC reference test model. Proposed modifications were evaluated experimentally and compared to unmodified 3D-HEVC and the 3D-HEVC codec adopted to compression of video acquired by camera with arbitrary locations as proposed in [1]. The latter appears to be more complex as its inter-view prediction is 44-fold slower than the interview prediction developed in this paper for circularly rectified video. Therefore, the technique seems to be an interesting proposal for applications within MPEG Immersive Video [10] where total bitrate for immersive video content can be reduced due to efficient exploitation of the inter-view redundancy.